\documentclass{mem}
\usepackage{natbib}\usepackage{txfonts}\usepackage{balance}
\usepackage{graphicx}
\usepackage{epstopdf}
\usepackage[a4paper]{hyperref}
\idline{?}{0}
\begin{document}
\title{Small-scale structure and dynamics of the chromospheric magnetic field}

\author{S. \, Wedemeyer-B\"ohm\inst{1,2}} 

\institute{
Institute of Theoretical Astrophysics, University of Oslo,
  P.O. Box 1029 Blindern, N-0315 Oslo, Norway
  \and
  Center of Mathematics for Applications (CMA), University of Oslo,
  Box 1053 Blindern, NÐ0316 Oslo, Norway
\email{sven.wedemeyer-bohm@astro.uio.no}
}

\authorrunning{Wedemeyer-B\"ohm}

\titlerunning{Small-scale structure and dynamics}

\abstract{
Recent advances in observational performance and numerical simulations have 
revolutionised our understanding of the solar chromosphere. 
This concerns in particular the structure and dynamics on small spatial and 
temporal scales. 
As a result, the picture of the solar chromosphere changed from an idealised 
static and plane-parallel stratification to a complex compound of intermittent 
domains, which are dynamically coupled to the layers below and above. 
In this picture, the chromosphere in a stricter sense is associated with the 
typical fibrillar structure shaped by magnetic fields like it is known from 
images taken in the H$\alpha$ line core. 
In internetwork regions below this layer, there exists a domain with  
propagating shock waves and weak magnetic fields, which both probably  
interact with the overlying large scale field. 
The existence of such a sub-canopy domain certainly depends on the properties 
of the overlying field. 
Details of the structure of the lower atmosphere can therefore be expected to 
vary significantly from location to location. 
Here, high-resolution observations, which were obtained with the CRISP filter 
at the Swedish Solar Telescope, are used to derive qualitative constraints for 
the atmospheric structure of quiet Sun regions.  
}
\maketitle

\section{Introduction}

\begin{table*}[t!]
\caption{CRISP data sets used in this study, scanning through the \ion{Ca}{II}\,854.2 line in June 2008.}
\vspace*{-6mm}
\begin{center}
\begin{tabular}{clccccccc}
\hline
data&
target&
date&
location&
\hspace*{-2mm}$\lambda$&
$(\Delta \lambda)_\mathrm{core}$& 
$(\Delta \lambda)_\mathrm{wing}$&
$\Delta{t}$& 
duration\\
set&&(June)&($\mu$)&\hspace*{-2mm}[pm]&[pm]&[pm]&[s]&[min]\\
\hline
1&coronal hole, quiet&13$^\mathrm{th}$& 0.92&\hspace*{-2mm} -92.0 -- 19.4 \hspace*{-2mm}&4.8& 4.8& 9&\hspace*{-1mm}33\\
2&coronal hole, quiet&15$^\mathrm{th}$& 0.99&\hspace*{-2mm}-193.9 -- 193.9\hspace*{-2mm}&9.7&19.4&11&\hspace*{-1mm}53\\
3&AR10998 (decayed)       &14$^\mathrm{th}$& 0.98&\hspace*{-2mm}-193.9 -- 193.9\hspace*{-2mm}&9.7&19.4&11&\hspace*{-1mm}41\\
4&AR10998 (pore)          &12$^\mathrm{th}$& 0.81&\hspace*{-2mm} -48.0 -- 48.0 \hspace*{-2mm}&9.7&19.4&11&\hspace*{-1mm}41\\
\hline
\end{tabular}
\end{center}
\label{wedemeyer-tab:datasets}
\end{table*}

Exploring the fine-structure of the solar chromosphere requires 
the combination of high spectral, temporal, and spatial 
resolution. 
In this respect, new instruments like 
the Interferometric BIdimensional Spectrometer 
\citep[IBIS, ][]{2006SoPh..236..415C} at the Dunn Solar Telescope 
and the CRISP filter \citep{2008ApJ...689L..69S} at the Swedish Solar Telescope (SST)
produced remarkable progress in our understanding of this atmospheric layer. 
\citet{2008A&A...480..515C} and \citet{2009A&A...494..269V} report that 
intensity maps obtained with IBIS in the \ion{Ca}{II}\,854.2\,nm line core 
show a fibrilar structure as it was known from H$\alpha$ observations. 
It implies that magnetic fields are an integral component even of 
quiet Sun regions. 
Quantitative measurements of the magnetic field above the photosphere 
are therefore essential for a better understanding of the structure and 
dynamics of the quiet Sun atmosphere. 
Despite the large progress in instrumentation, such measurements are still very 
difficult as demonstrated by 
\citet{2009ApJ...706..148W} 
and 
\citet[][ see this volume]{2010memsait_sacpeak_jaime} 
using IBIS and CRISP, respectively.  
In particular the presumably weak magnetic fields above the internetwork photosphere are 
difficult to measure. 
Until even better instruments become available, we are mostly bound to derive qualitative 
constraints based on high-resolution intensity map series in close comparison with 
current numerical simulations. 

In this paper, such qualitative (but still preliminary) constraints are derived 
from the analysis of CRISP data. 
Observations of regions in a coronal hole with apparently very weak magnetic 
field are compared to a peculiar region outside the hole that remained after 
the decay of an active region.

\section{Observations}
\label{wedemeyer-sec:observ}

The CRISP instrument is used to scan through the \ion{Ca}{II}\,IR line at $\lambda = 854.2$\,nm. 
See Table~\ref{wedemeyer-tab:datasets} for an overview over the data sets. 
All data sets are complemented with wide-band (WB) images.  
The MOMFBD code by \citet{2005SoPh..228..191V} was used for 
image restoration. 
The data sets were obtained at different days at positions 
close to disc-centre. 
Sets~1 (June 13th, 2008) and 2 (June 15th, 2008) were taken in quiet 
Sun regions in a coronal hole. 
Data set~3 (June 14th, 2008) shows a very peculiar region  
in the remaining of a decaying active region (AR10998).   
Two days earlier (June 12th, 2008) there a pore was visible at that location 
(set 4). 
The data sets are complementary concerning the topology of the magnetic field. 
The cell in set~3 is covered by a presumably horizontally 
aligned field, whereas the quiet Sun sets 1 and 2 show a less dense 
coverage with magnetic field and probably a more vertically aligned field.

\section{Inside a coronal hole}
\label{wedemeyer-sec:corhole}

\subsection{Atmospheric structure} 

\begin{figure*}
\centering 
\includegraphics[width=\textwidth]{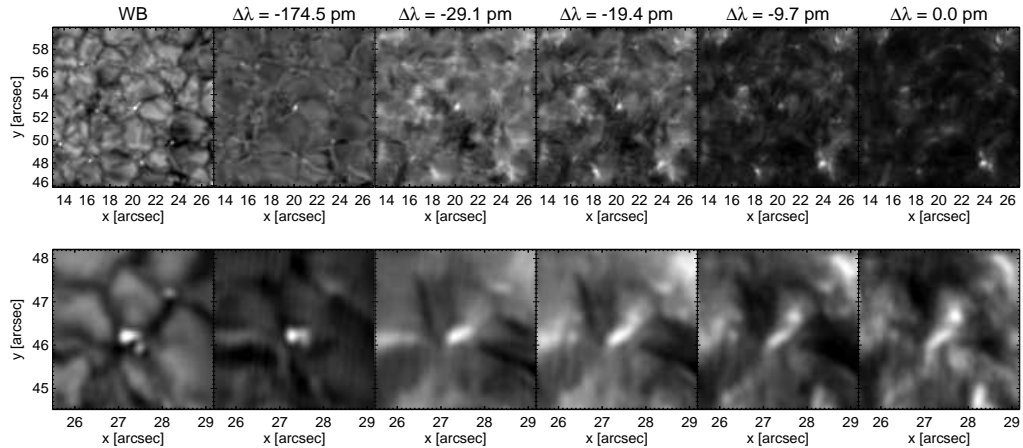}
\caption{\footnotesize
CRISP scans through the Ca line in data set~2 for a close-up region 
(top) and strongly magnified (bottom).  
The WB maps and a few wavelength steps are shown (left to right). 
}
\label{wedemeyer-fig:datasethole}
\end{figure*}

The CRISP WB images essentially map the continuum intensity formed in 
the low photosphere. 
In data sets 1 and 2, these maps show normal quiet Sun 
granulation with a few bright grains in the intergranular lanes 
(Fig.~\ref{wedemeyer-fig:datasethole}). 
These grains are known to be associated with magnetic field structures. 
By stepping through the Ca line towards the core, the maps in principle show atmospheric 
layers at increasing height. 
The line wing maps at 854.147\,nm  are clearly dominated by the 
reversed granulation pattern, which is formed in the middle photosphere. 
The most pronounced features can still be identified 
in the maps close to the line core. 
The core maps exhibit only very few fibrils (which would be part of the magnetic 
chromosphere in a stricter sense) and a faint mesh-like pattern in-between. 
The latter is most likely produced by the interaction of shock waves. 
Such a domain was called ``clapotisphere'' by \citet{1995ESASP.376a.151R}
and alternatively ``fluctosphere'' by \citep{2008IAUS..247...66W}\label{wedemeyer-sec:deffluctosphere}. 
The reversed granulation and fluctospheric pattern blend smoothly as one steps  
through the wavelength. 
Most of the bright structures that are obvious in the line core maps  
can only be discerned up to 0.24\,nm into the wing.

\subsection{Internetwork bright points}
\label{wedemeyer-sec:qsbp}

Most bright points (BPs) can be followed through the Ca line towards clear 
counterparts in the line core. 
Their positions are not very different in the WB and the line core maps.  
The size is generally more extended in the Ca line core compared to the WB. 
In the idealised ``flux tube'' picture, this would imply a predominantly 
vertically orientated funnel, which extends with height in response to the 
decreasing thermal pressure of the ambient medium. 
However, the scene in the maps in and close to the Ca line core leaves no doubt that 
this picture is oversimplified. 
For some WB BPs, there is no clear counterpart in the maps closer to the line core
but only a faint and diffuse feature. 
Vice versa, some roundish brightenings, which are present throughout the CRISP scans, 
cannot unambiguously be identified in the WB.
BPs often appear in small groups in the WB maps. 
From this alone, the change of the magnetic field with height cannot conclusively 
be deferred, demonstrating the need for direct measurements of the magnetic field.

\subsection{Propagating shock fronts} 
\label{wedemeyer-sec:set1_shockwaves}

Propagating shock fronts seem to be an ubiquitous phenomenon in the Ca line core maps.  
There are examples of bright grains that evolve into granule-size 
bright rings before fading away. 
The estimated lifetime of such events is of the order of 100\,s. 
Horizontal shock expansion speeds seem to be of the order of 8 -- 10 km\,s$^{-1}$. 
The intensity signature reminds of the one seen in the fluctosphere of 3D simulations  
as described, e.g., by \citet[][ see Fig.~15 therein]{2009SSRv..144..317W}. 
The ubiquitous shock waves in the simulations can be seen as 
expanding rings of enhanced intensity in the \ion{Ca}{II} line cores. 
The strongest brightening occurs at sites where neighbouring shock fronts collide, 
compress and heat the gas to high temperatures.  
These collision zones are then visible as thin elongated threads and/or small 
grains. 
The CRISP data discussed here has the sufficient spatial, temporal, {\em and} spectral 
resolution to reveal these details. 
As demonstrated by \citet{2008IAUS..247...66W}, limited instrumental possibilities
can easily smear out most of the faint fine-structure so that only the brightest 
features are detected. 
That is probably why so far mostly short-lived Ca~grains were observed 
\citep[e.g.,][]{1991SoPh..134...15R, 1993A&A...274..584K}.

\subsection{Calcium swirls }
\label{wedemeyer-sec:swirl}

The line core maps of data sets~1 and 2 exhibit small-scale swirl events 
\citep[][ see also the movie in the online material]{2009A&A...507L...9W}.
They consist of ring fragments and/or spiral arms that 
appear to rotate around small groups of bright points in the centre. 
The diameters of the rotating regions are typically $\sim 2$\,\arcsec. 
The width of the features is of the order of only 0\,\farcs2, which is close to 
the effective spatial resolution of the observations. 
The line core is blue-shifted at positions in the dark fragments, corresponding to 
upward velocities of the order of 2 to $\sim 5$\,km$/$s. 
The blue-shift often increases along the dark spiral arms outwards,   
reaching velocities of up to $\sim 7$\,km$/$s. 
It seems as if plasma is accelerated and ejected. 
This phenomenon is most likely caused by plasma spiraling upwards in a funnel-like 
magnetic field structure. 
The motion of the related photospheric BPs suggests that the magnetic footpoints 
of these structures are buffeted by the convective flows, resulting in the 
twisting and braiding of the field above. 
This mechanism would have implications for coronal heating 
\citep{1988ApJ...330..474P}. 
Current 3D MHD simulations with \mbox{CO$^5$BOLD} show ring fragments in magnetic 
funnels, which probably are related to swirls. 

\section{A peculiar region outside a coronal hole}
\label{wedemeyer-sec:cell}

\begin{figure}
\centering 
\includegraphics[width=6.5cm]{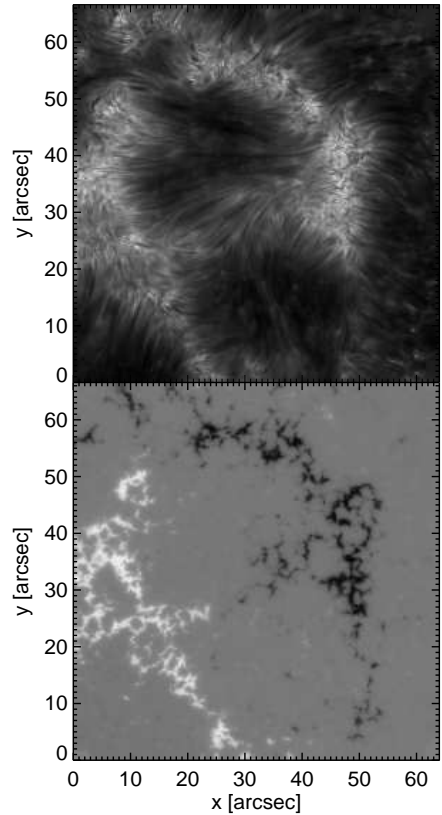}
\vspace*{-5mm}
\caption{\footnotesize
Inner part of the FOV in data set~3 featuring a prominent   
region. 
\textit{Top:} \ion{Ca}{II}\,intensity map at $\lambda = 854.2$\,nm
(average line core position);  
\textit{bottom:} magnetogram taken 5\,min after the intensity map. 
The left part of the cell has positive Stokes~V (white), whereas the right side 
has negative (black). 
}
\label{wedemeyer-fig:datasetcell}
\end{figure}

Data sets~3 and 4 were recorded in a decaying active region. 
The prominent pore in data set 4 has vanished and is no longer visible in 
set~3 just two days later.  
Instead we see a well outlined region that resembles a network cell but probably 
has a stronger magnetic field.  
The two halves have opposite polarity as can be seen from the magnetogram in 
Fig.~\ref{wedemeyer-fig:datasetcell}  (recorded five minutes after the Ca line scan).
The magnetic field lines connecting the two polarities are therefore neatly aligned, 
resulting in a magnetic field ``canopy'', which is visible in form of  
fibrils in the Ca line core maps. 
The Stokes~V signal in the interior is predominantly weakly negative except for 
a small number of ``islands'' with positive signal, which are more frequent towards 
the left-hand part.  
The granulation in the cell interior appears to be normal.

\subsection{Shock waves}
\label{wedemeyer-sec:shockwaves}

\begin{figure}
\centering 
\includegraphics{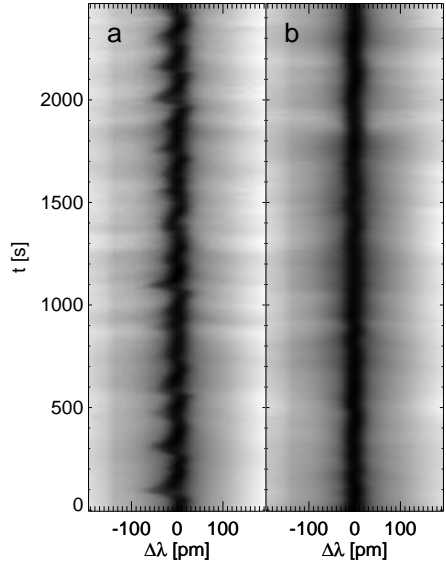}
\vspace*{-2mm}
\caption{\footnotesize 
Spectra of the \ion{Ca}{II}\,854 line as function of time for two pixels in data set~3. 
\textbf{a)}~outside the cell (lower right corner of the FOV, [60\,\farcs{}4, 8\,\farcs{}0]);  
\textbf{b)}~inside the cell at [23\,\farcs{}9, 44\,\farcs{}9].  
Coordinates refer to Fig.~\ref{wedemeyer-fig:datasetcell}. 
}
\label{wedemeyer-fig:spectrumtwo}
\end{figure}

In Fig.~\ref{wedemeyer-fig:spectrumtwo}a, the spectrum of the Ca line is shown as function 
of time for a position outside the peculiar cell. 
It shows a clear signature of propagating shock waves with characteristic 
blue and red shifts of the line core. 
The quiet Sun data sets 1 and 2 show an even more pronounced shock wave
activity. 
In contrast, the spectrum taken at a position in the cell interior shows essentially
no shock wave signature. 
There is only a small intensity excess moving through the spectrum with time, which can be 
interpreted as the onset of shock formation without a sign of a propagating 
shock wave afterwards.

As the granulation (low photosphere) and the reversed granulation (middle 
photosphere) look ``normal'', it is likely that acoustic waves are excited 
in the lower atmosphere in the cell interior. 
The absence of shock wave signatures is therefore probably caused by  
the magnetic field in the layers above. 
A horizontal magnetic field, as it is seen in form of Ca line core fibrils, 
may hinder the steeping of the initial acoustic waves into shock waves. 
It possibly includes dissipation of wave energy, (partial) wave mode conversion, 
and the suppression of downflowing material which may contribute to the shock 
formation process.  
Based on the results from numerical simulations the shocks are typically formed 
at heights between 700\,km and 1000\,km. 
The normal appearance of the reversed granulation pattern together with the absence of 
shocks could be interpreted such that the fibrilar magnetic field spanning the region is
possibly located higher than 400\,km but well below 1000\,km. 
This does not exclude that the field covers the height range above, too.

\subsection{Magnetic field topology above internetwork bright points} 
\label{wedemeyer-sec:magtopointer}

\begin{figure*}
\centering 
\includegraphics[width=\textwidth]{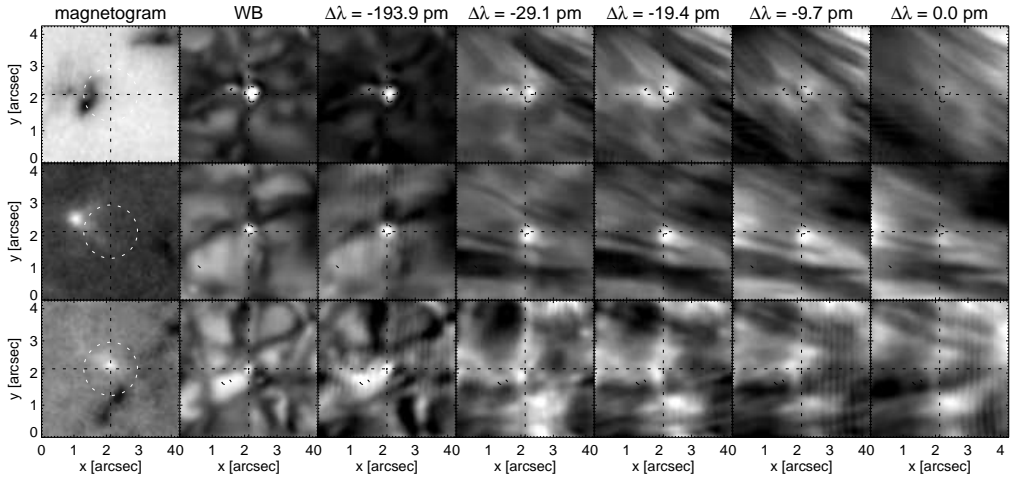} 
\vspace*{-5mm}
\caption{\footnotesize
Bright points in data set~3. Each row shows a different example. 
The leftmost column shows the wideband image, the next in the blue wing of the 
Ca line, and the others on the blue side of the line core.  
The dotted contour mark the bright points as seen in the wideband. }
\label{wedemeyer-fig:brightpointsjun14}
\end{figure*}

There are a number of photospheric BPs in the cell interior, which  
stay in place for quite some time. 
They are visible in the WB maps and throughout the whole 
Ca line scans.  
A few examples are shown in Fig.~\ref{wedemeyer-fig:brightpointsjun14}. 
The upper row is for a BP, which is located close to the cell boundary. 
The leftmost panel shows the magnetogram for that region.  
As the magnetogram was recorded five minutes after the Ca line scan, small differences
due to evolution and advection are to be expected. 
Nevertheless, the magnetogram shows that this particular BP is connected 
to negative Stokes~V. 
Most of the other BPs, especially those further away 
from the left cell boundary, are instead connected to positive Stokes~V.

Like in the WB images, the BPs are seen as roundish bright features in the Ca maps 
when stepping through wavelength towards the line core. 
In many cases, this holds up to 20 to 30\,pm from the central wavelength.  
Closer to or at the line core, fibrils become visible. 
In most cases, the BP appears as faint knot on a fibril.  
There are examples for which a few secondary knots appear next to the main BP. 
These knots can be found either on the same fibril as the main BP or at neighbouring 
fibrils. 
Finally, there are examples for which more extended elongated bright features appear next to 
the main BP. 
These can even be orientated perpendicular to the fibrils. 
A possible explanation is that the bright knots on the fibrils are connection points 
between the BP fields and the overlying ``canopy'' field, which itself is anchored in the magnetic 
network boundaries. 
However, it is not possible to deduce the magnetic field topology from the intensity maps 
alone. 
The knots could therefore just be a line-of-sight effect, with the BPs shining through the 
fibrils. 

\subsection{Bright blobs moving along fibrils}
\label{wedemeyer-sec:movingblobs}

Individual bright blobs appear to move along the fibrils in the Ca line core maps, 
thus following the magnetic field across the network cell.   
There are examples for propagation from W to E but also from E to W. 
The blobs are often comparable in size with the fibrils but can also be somewhat 
larger.
Sometimes there are chains of bright blobs that move next to each other along 
neighbouring fibrils. 
Data set~4 even exhibits pulses that seem to run from the 
magnetic field structure in the middle of the FOV radially outwards, following 
the fibrils. 

\begin{figure*}
\centering 
\includegraphics[width=\textwidth]{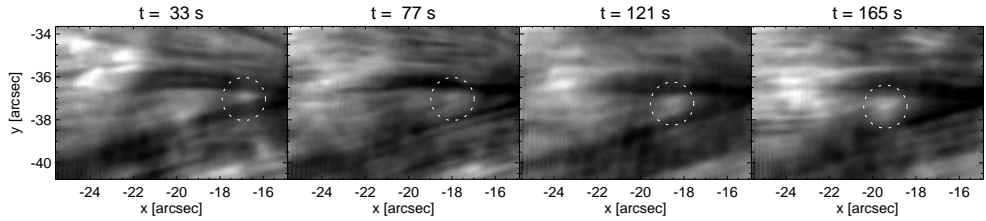} 
\vspace*{-5mm}
\caption{\footnotesize 
	Close-ups of Ca line core maps from data set~3 for the Ca line wing.
	The panels from left to right show the temporal evolution.
	The dotted circles mark a bright blob that propagates parallel to the fibrils. 
}
\label{wedemeyer-fig:movblob}
\end{figure*}

The panels in Fig.~\ref{wedemeyer-fig:movblob} show close-ups of a blob for four time steps 
in the Ca line core. 
The feature is not visible in the line wing maps, implying that this phenomenon 
truly originates in a layer above the photosphere. 
The blob can be tracked over $\sim 11$\,min before it vanishes close to 
other the cell boundary. 
The apparent propagation speed is of the order of 4 to 8\,km\,s$^{-1}$ with a 
possible (and yet to be confirmed) modulation with a period of 7\,min. 

A likely interpretation is that waves are excited at the photospheric footpoints 
of the field structure in the magnetic cell boundary and then move along the field lines.  
Another possibility is that a upward propagating wave front reaches the canopy fibrils  
from below at consecutive times, producing the impression of a moving blob.  
Interaction with shock waves would be a potential excitation mechanism for the observed lateral 
swaying of fibrils.

\subsection{Diffuse features}
\label{wedemeyer-sec:diffeat}

Finally, there are fast propagating features of slightly enhanced intensity that do 
not follow the magnetic field associated with the fibrils. 
They have larger extent than the BPs and blobs discussed above, appear more diffuse 
and propagate fast. 
A possible explanation is that they are wave fonts being ``scattered'' at the lower 
boundary of the magnetic ``canopy'' field. 
However, a more thorough analysis of such events is required. 

\section{Discussion}
\label{wedemeyer-sec:discus}

\paragraph{Coronal hole data:} 
The CRISP images taken in the coronal hole (Sect.~\ref{wedemeyer-sec:corhole}) show only 
only few fibrils. 
The same is found in corresponding H$\alpha$ images, which were taken in the same 
FOV as data set~2 directly after the Ca series. 
In contrast, the Ca line core images in set~4  outside the coronal hole 
(even in a quiet region away from the pore) are characterised by fibrils. 
The latter agrees with IBIS observations by \citet{2008A&A...480..515C} who report 
a high coverage of fibrils in line core maps even in quiet Sun regions. 
The magnetic field configuration in- and outside coronal holes is obviously very 
different. 
It has direct consequences for the magnetic field structure on smaller scales 
in the layer above the photosphere which is mapped in the 
\ion{Ca}{II}\,854.2\,nm line core. 
It is likely that the chromosphere in coronal holes has no pronounced canopy field 
but rather contains "open" and more vertically orientated field lines.  
At least it would explain why swirls were only found in the coronal hole data away from 
fibrils. 
Horizontal or at least more inclined fields, which would show up as fibrils, are likely 
to hinder the rotating motion associated with swirls. 

Next to the fibrils, the CRISP line core maps contain a short-lived 
mesh-like pattern with bright grains at the vertices.  
It is similar to the pattern observed by \citet{2006A&A...459L...9W} with 
a particularly narrow Lyot filter (FWHM 30\,pm) at the 
German Vacuum Tower Telescope. 
This pattern is very likely the intensity signature of a ``fluctosphere'' 
\citep[see Sect.~\ref{wedemeyer-sec:deffluctosphere} here and Fig.~15 in][]{2009SSRv..144..317W} 
as it is present in numerical 3D simulations that include no 
or only weak magnetic fields 
\citep[e.g.,][]{skartlien00c,2004A&A...414.1121W,2008ApJ...679..871M}.

\paragraph{The peculiar region:}  
In contrast to the coronal hole data, the peculiar region analysed in 
Sect.~\ref{wedemeyer-sec:cell} has a strong magnetic ``canopy'' field, which is visible 
as fibrils spanning across the cell. 
Most of them are far from static but sway laterally with a range of different 
apparent periods. 
The fibrils observed in the Ca line core are not necessarily the same as sampled 
in the H$\alpha$ line. 
It could well be so that there exist several layers of fibrils with the lower laying
ones visible in the Ca line core and those above visible in the H$\alpha$ line.  

The normal appearance of reversed granulation and the absence of shock 
signatures would constrain the lower boundary of such a magnetic canopy to a 
layer between $\sim 400$\,km and $\sim 700 - 1000$\,km. 
\citep[These numbers are based on numerical simulations by, e.g.,][]{2008ApJ...679..871M, 
 2004A&A...414.1121W}. 
A possible explanation for a low boundary is that the fibrils sink down in the atmosphere
due to ``mass-loading'', i.e., mass injected into the fibrils at their footpoints. 
The features discussed in Sect.~\ref{wedemeyer-sec:movingblobs} could indeed indicate 
parcels of gas moving along the fibrils.
The diffuse features in Sect.~\ref{wedemeyer-sec:diffeat} might be ``failed shocks'' -- 
waves approaching a low canopy from below and getting scattered and damped 
before they can steepen into the saw-tooth profile so prominently seen in the 
the coronal hole data. 

An alternative explanation for the absence of shocks could be that the overlying 
magnetic canopy forms a cavity that favors standing waves over propagating waves. 

\section{Conclusion}
\label{wedemeyer-sec:conc}

During the last years, our picture of the lower solar atmosphere changed from a 
static plane-parallel stratification to a very complex, coupled 
compound of dynamic domains 
\citep[see the recent reviews by, e.g.,][]{2001ASPC..223..131S, 2006ASPC..354..259J, 
2007ASPC..368...27R, 2009SSRv..144..317W}. 
Outside coronal holes, the magnetic field above the photosphere often appears 
to consist of a large-scale canopy and an enclosed sub-canopy domain with presumably 
weaker field. 
The transition between these components is certainly continuous. 
The sub-canopy domain may host a ``fluctosphere''/``clapotisphere'' produced by shock 
waves.
The existence of such a domain, however, critically depends on the properties of the 
overlying ``canopy'' field. 
The example discussed in Sect.~\ref{wedemeyer-sec:cell} suggests that a canopy at 
low height could possibly suppress a fluctosphere. 
The atmospheric structure might therefore vary significantly between different 
quiet Sun regions. 
The differences are particularly large compared to regions inside 
coronal holes, which do not seem to possess a pronounced horizontal 
field component as in a canopy. 
The absence of such a component could make the existence of a fluctosphere more likely in these regions.

In regions with a magnetic canopy, upwards propagating shock waves have to 
meet the overlying magnetic field at some point but would already interact  
with the (supposedly weaker) magnetic field in the sub-canopy domain on their way up  
\citep{2009arXiv0904.2026S}.
The details of such interactions still need further investigation.  
It certainly includes mode conversion and wave guiding and may have   
implications for the energy transport in the quiet Sun on small spatial scales.


\begin{acknowledgements}
The author thanks the organisers of the 25th SacPeak Workshop for the invitation
and all colleagues from Oslo who helped with the observations and the data 
reduction, in particular Luc Rouppe van der Voort. 
This work was supported by the Research Council of Norway, grant 
170935/V30, and through a Marie Curie Intra-European Fellowship of the European 
Commission (FP6-2005-Mobility-5, no. 042049). 
\end{acknowledgements}

\bibliographystyle{aa}

\end{document}